\RequirePackage{fix-cm}
\documentclass[smallextended]{svjour3}
\smartqed
\usepackage{graphicx}

\usepackage{amsmath}
\usepackage{amssymb}
\usepackage{braket}
\usepackage{color}
\usepackage{kbordermatrix}

\newtheorem{thm}{Theorem}

\begin{document}

\title{Limit distribution of a continuous-time quantum walk with a spatially 2-periodic Hamiltonian
}
\subtitle{}

\titlerunning{Limit distribution of a continuous-time quantum walk}

\author{Takuya Machida}

\authorrunning{T.~Machida}

\institute{%
T.~Machida \at
              College of Industrial Technology, Nihon University, Narashino, Chiba 275-8576, Japan\\
              \email{machida.takuya@nihon-u.ac.jp}\\
}

\date{}

\maketitle

\begin{abstract}
Focusing on a continuous-time quantum walk on $\mathbb{Z}=\left\{0,\pm 1,\pm 2,\ldots\right\}$, we analyze a probability distribution with which the quantum walker is observed at a position.
The walker launches off at a localized state and its system is operated by a spatially periodic Hamiltonian.
As a result, we see an asymmetric probability distribution.
To catch a long-time behavior, we also try to find a long-time limit theorem and realize that the limit distribution holds a symmetric density function.

\end{abstract}

\section{Introduction}
Quantum walks are considered as quantum analogues of random walks.
Motivated by quantum computation, continuous-time quantum walks were introduced in 2002~\cite{ChildsFarhiGutmann2002}.
They are defined by discrete-space Schr\"{o}dinger equations and have been studied in Mathematics, Physics, and quantum information.
In Mathematics, one of the aims is to get long-time limit theorems for quantum walks because we understand from the limit theorems how the quantum walker behaves after long-time evolution.
As a result, some interesting properties of the quantum walks are discovered and they can be applied to computer science, for instance, quantum algorithms~\cite{Venegas-Andraca2012}.
Continuous-time quantum walks are modeled by adjacency matrices of the graphs, which are equivalent to Hamiltonians of the quantum walks.
While a lot of limit theorems have been reported in discrete-time quantum walks, we see some limit theorems in continuous-time quantum walks.
In Refs.~\cite{Konno2005b,Gottlieb2005}, a limit distribution of a continuous-time quantum walk on $\mathbb{Z}=\left\{0,\pm1,\pm2,\ldots\right\}$ was reported, and its density function was given by $1/\pi\sqrt{1-x^2}$.
Konno~\cite{Konno2006} demonstrated a limit theorem for continuous-time quantum walks on trees and the limit density function held $x^2/\pi\sqrt{4-x^2}$.
In Ref.~\cite{Monvel2022}, only the convergence for the position of continuous-time quantum walkers on general $Z^d$-periodic graphs was proved in Theorem 5.5 of the paper.
We also aim at a limit distribution of a continuous-time quantum walk.
The quantum walker moves on $\mathbb{Z}$ and its system gets updated by a spatially periodic Hamiltonian.

The rest of this paper has five sections.
We start off with the definition of a continuous-time quantum walk in Sect.~\ref{sec:definition} and find its probability amplitude in integral forms in Sect.~\ref{sec:probability_amplitude}.
The limit distribution is demonstrated by Fourier analysis in Sect.~\ref{sec:limit_distribution}, compared to numerical experiments.
Section~\ref{sec:summary} is assigned for a short discussion.
The results in a special case will be reported without proving them in the appendix.

\section{Definition of a continuous-time quantum walk}
\label{sec:definition}
\begin{figure}[h]
 \begin{center}
  \includegraphics[scale=0.6]{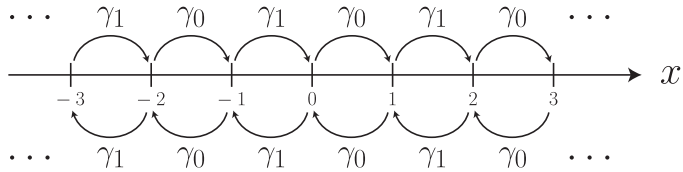}
  \caption{The Hamiltonian given by Eqs.~\eqref{eq:time_ev_even} and \eqref{eq:time_ev_odd} is spatially 2-periodic.}
  \label{fig:1}
 \end{center}
\end{figure}

The system of continuous-time quantum walk at time $t\,(\,\geq 0)$ is described by probability amplitude $\left\{\psi_t(x)\in\mathbb{C} : x\in\mathbb{Z}\right\}$, where $\mathbb{C}$ is the set of complex numbers.
The quantum walker launches off at a localized initial state,
\begin{equation}
 \psi_0(x)=\left\{\begin{array}{ll}
  1 & (x=0)\\
	    0 & (x\neq 0)
	   \end{array}\right..\label{eq:initial}
\end{equation}
With two real numbers $\gamma_0$ and $\gamma_1$, the probability amplitude at time $t$ gets updated in a Schr\"{o}dinger equation,
\begin{align}
 i\,\frac{d}{dt}\psi_t(2n)=& \gamma_1\,\psi_t(2n-1)+\gamma_0\,\psi_t(2n+1),\label{eq:time_ev_even}\\
 i\,\frac{d}{dt}\psi_t(2n+1)=& \gamma_0\,\psi_t(2n)+\gamma_1\,\psi_t(2n+2)\label{eq:time_ev_odd},
\end{align}
where $n\in\mathbb{Z}$ and $i$ denotes the imaginary unit.
The Schr\"{o}dinger equation is equivalently depicted in a matrix form,
\begin{equation}
i\,\frac{d}{dt}
\kbordermatrix{
& \\
& \vdots\\
&  \psi_t(-3)\\
&  \psi_t(-2)\\
& \psi_t(-1)\\
& \psi_t(0)\\
& \psi_t(1)\\
& \psi_t(2)\\
& \psi_t(3)\\
& \vdots
}
=
\kbordermatrix{
   & \cdots & -3 & -2 & -1 & ~0 & ~1 & ~2 & ~3 & \cdots \\
\vdots & \ddots & \cdot  &  \cdot &  \cdot &  \cdot & \cdot & \cdot & \cdot & \ldots \\
-3 & \cdots &  0 &  \gamma_1 &  0 & 0 &  0 &  0  & 0  & \cdots \\
-2 & \cdots &  \gamma_1 &  0 &  \gamma_0 & 0 &  0 &  0  & 0  &  \cdots \\
-1 & \cdots &  0 &  \gamma_0 &  0 & \gamma_1 &  0 &  0  & 0  & \cdots \\
~0  &  \cdots & 0 &  0 &  \gamma_1 & 0 & \gamma_0 & 0 & 0  & \cdots \\
~1  &  \cdots & 0 &  0 &  0 & \gamma_0 & 0 & \gamma_1 & 0  & \cdots \\
~2  &  \cdots & 0 &  0 &  0 & 0 & \gamma_1 & 0 & \gamma_0  & \cdots \\
~3  &  \cdots & 0 &  0 &  0 & 0 & 0 & \gamma_0 & 0  & \cdots \\
\vdots & \cdots & \cdot  &  \cdot &  \cdot &  \cdot & \cdot & \cdot & \cdot & \ddots 
}
\kbordermatrix{
& \\
& \vdots\\
&  \psi_t(-3)\\
&  \psi_t(-2)\\
& \psi_t(-1)\\
& \psi_t(0)\\
& \psi_t(1)\\
& \psi_t(2)\\
& \psi_t(3)\\
& \vdots
}
.
\end{equation}
The Hamiltonian
\begin{equation}
\kbordermatrix{
   & \cdots & -3 & -2 & -1 & ~0 & ~1 & ~2 & ~3 & \cdots \\
\vdots & \ddots & \cdot  &  \cdot &  \cdot &  \cdot & \cdot & \cdot & \cdot & \ldots \\
-3 & \cdots &  0 &  \gamma_1 &  0 & 0 &  0 &  0  & 0  & \cdots \\
-2 & \cdots &  \gamma_1 &  0 &  \gamma_0 & 0 &  0 &  0  & 0  &  \cdots \\
-1 & \cdots &  0 &  \gamma_0 &  0 & \gamma_1 &  0 &  0  & 0  & \cdots \\
~0  &  \cdots & 0 &  0 &  \gamma_1 & 0 & \gamma_0 & 0 & 0  & \cdots \\
~1  &  \cdots & 0 &  0 &  0 & \gamma_0 & 0 & \gamma_1 & 0  & \cdots \\
~2  &  \cdots & 0 &  0 &  0 & 0 & \gamma_1 & 0 & \gamma_0  & \cdots \\
~3  &  \cdots & 0 &  0 &  0 & 0 & 0 & \gamma_0 & 0  & \cdots \\
\vdots & \cdots & \cdot  &  \cdot &  \cdot &  \cdot & \cdot & \cdot & \cdot & \ddots 
},
\end{equation}
is, therefore, spatially 2-periodic, as shown in Fig.~\ref{fig:1}.

\noindent
We assume that the values of parameters $\gamma_0$ and $\gamma_1$ are not zero (i.e. $\gamma_0, \gamma_1\neq 0$), and $|\gamma_0|\neq |\gamma_1|$.
We will discuss about the special case $|\gamma_0|=|\gamma_1|$ in the appendix.
Let $X_t$ be the position of the quantum walker at time $t$.
The walker is observed at position $x$ at time $t$ with probability
\begin{equation}
 \mathbb{P}(X_t=x)=\bigl|\psi_t(x)\bigr|^2.
\end{equation}

Let us introduce the Fourier transforms of the amplitude, represented by $\hat\psi_{0,t}(k)$ and $\hat\psi_{1,t}(k)\,(k\in [-\pi, \pi))$,
\begin{align}
 \hat\psi_{0,t}(k)=& \sum_{n\in\mathbb{Z}}e^{-ik\cdot 2n}\psi_t(2n),\\
 \hat\psi_{1,t}(k)=& \sum_{n\in\mathbb{Z}}e^{-ik\cdot (2n+1)}\psi_t(2n+1).
\end{align}
By the inverse Fourier transform, the Fourier transforms get back to the amplitude,
\begin{align}
 \psi_t(2n)=& \int_{-\pi}^{\pi} e^{ik\cdot 2n}\hat\psi_{0,t}(k)\,\frac{dk}{2\pi},\\
 \psi_t(2n+1)=& \int_{-\pi}^{\pi} e^{ik\cdot (2n+1)}\hat\psi_{1,t}(k)\,\frac{dk}{2\pi}.
\end{align}
Equations~\eqref{eq:time_ev_even} and \eqref{eq:time_ev_odd} bring the time evolution of the Fourier transforms, 
\begin{align}
 i\,\frac{d}{dt}\hat\psi_{0,t}(k)=& (\gamma_1e^{-ik}+\gamma_0e^{ik})\hat\psi_{1,t}(k),\label{eq:time_ev_f0}\\
 i\,\frac{d}{dt}\hat\psi_{1,t}(k)=& (\gamma_0e^{-ik}+\gamma_1e^{ik})\hat\psi_{0,t}(k),\label{eq:time_ev_f1}
\end{align}
and Eq.~\eqref{eq:initial} gives the initial conditions $\hat\psi_{0,0}(k)=1$ and $\hat\psi_{1,0}(k)=0$.

\section{Probability amplitude}
\label{sec:probability_amplitude}
We find the solution of Eqs.~\eqref{eq:time_ev_f0} and \eqref{eq:time_ev_f1},
\begin{align}
 \hat\psi_{0,t}(k)=& \frac{1}{2}\left(e^{i\sqrt{g(k)}\cdot t}+e^{-i\sqrt{g(k)}\cdot t}\right)
 = \cos\left(\sqrt{g(k)}\cdot t\right),\\
 \hat\psi_{1,t}(k)=& -\frac{\sqrt{g(k)}}{2I(k)}\left(e^{i\sqrt{g(k)}\cdot t}-e^{-i\sqrt{g(k)}\cdot t}\right)
 = -i\,\frac{\sqrt{g(k)}}{I(k)}\,\sin\left(\sqrt{g(k)}\cdot t\right),
\end{align} 
where
\begin{align}
 g(k)=& \,\gamma_0^2+\gamma_1^2+2\gamma_0\gamma_1\cos 2k,\\
 I(k)=& \,\gamma_0 e^{ik}+\gamma_1 e^{-ik}.
\end{align}
We should note $|I(k)|=\sqrt{g(k)}$.
Using the inverse Fourier transform, we get the probability amplitude at time $t$ in integral representations,
\begin{align}
 \psi_t(2n)
 =& \int_{0}^{\frac{\pi}{2}}\frac{2}{\pi}\cos(2nk)\cos\left(\sqrt{g(k)}\cdot t\right)\,dk,\\
 \psi_t(2n+1)
 =& -i\,\int_{0}^{\frac{\pi}{2}}\frac{2}{\pi}\Bigl\{\gamma_0\cos (2nk)+\gamma_1\cos \bigl(2(n+1)k\bigr)\Bigr\}\,\frac{\sin\left(\sqrt{g(k)}\cdot t\right)}{\sqrt{g(k)}}\,dk.
\end{align}
The amplitudes $\psi_t(-2n)$ and $\psi_t(2n)$ are symmetric at any time $t$, that is, $\psi_t(-2n)=\psi_t(2n)$.
On the other hand, we can not say $\psi_t(-(2n+1))=\psi_t(2n+1)$.

\section{Limit distribution}
\label{sec:limit_distribution}

As many limit distributions have been derived for a scaled position by time $t$ in quantum walks, one can claim a limit theorem for the continuous-time quantum walk. 
\begin{thm}
Let $\xi\in\left\{0,1\right\}$ be the subscription such that $|\gamma_\xi|=\min\left\{\,|\gamma_0|, |\gamma_1|\,\right\}$.
The quantum walker is supposed to localize at position $x=0$ at time $t=0$, that is, $\psi_0(0)=1$ and $\psi_0(x)=0\,(x\neq 0)$.
Then, for a real number $x$, we have
 \begin{equation}
   \lim_{t\to\infty}\mathbb{P}\left(\frac{X_t}{t}\leq x\right)=\int_{-\infty}^x \frac{1}{\pi\sqrt{4\gamma_\xi^2-y^2}}I_{(\,-2|\gamma_\xi|, 2|\gamma_\xi|\,)}(y)\,dy,
 \end{equation}
 where
 \begin{equation}
  I_{(\,-2\,|\gamma_\xi|, 2\,|\gamma_\xi|\,)}(x)
   =\left\{\begin{array}{cl}
     1&(\,-2\,|\gamma_\xi| < x < 2\,|\gamma_\xi|\,)\\
	    0&(\mbox{otherwise})
	   \end{array}\right..
 \end{equation}
 \label{th:limit}
\end{thm}

\begin{figure}[h]
\begin{center}
 \begin{minipage}{50mm}
  \begin{center}
   \includegraphics[scale=0.4]{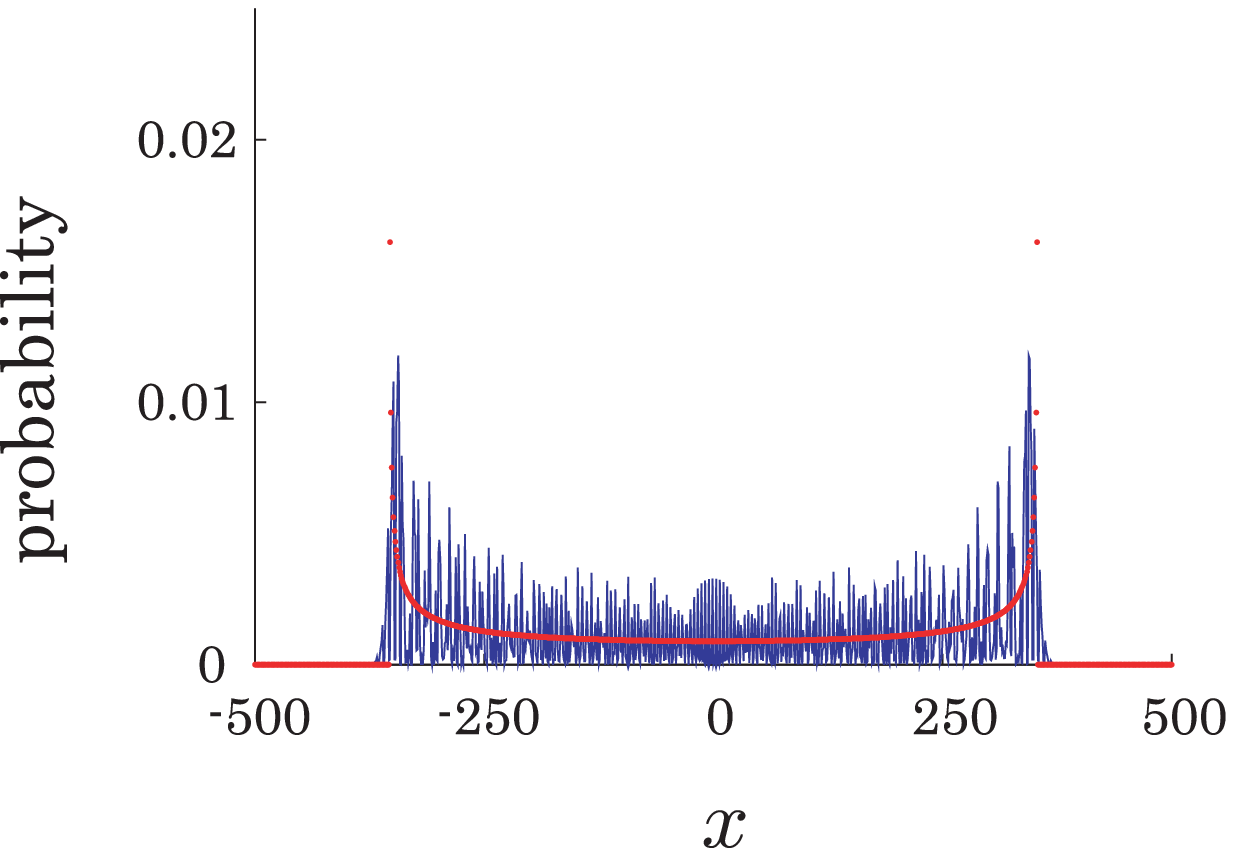}\\[2mm]
  (a) $\gamma_0=1/2\sqrt{2},\, \gamma_1=1/\sqrt{2}$
  \end{center}
 \end{minipage}\hspace{10mm}
 \begin{minipage}{50mm}
  \begin{center}
   \includegraphics[scale=0.4]{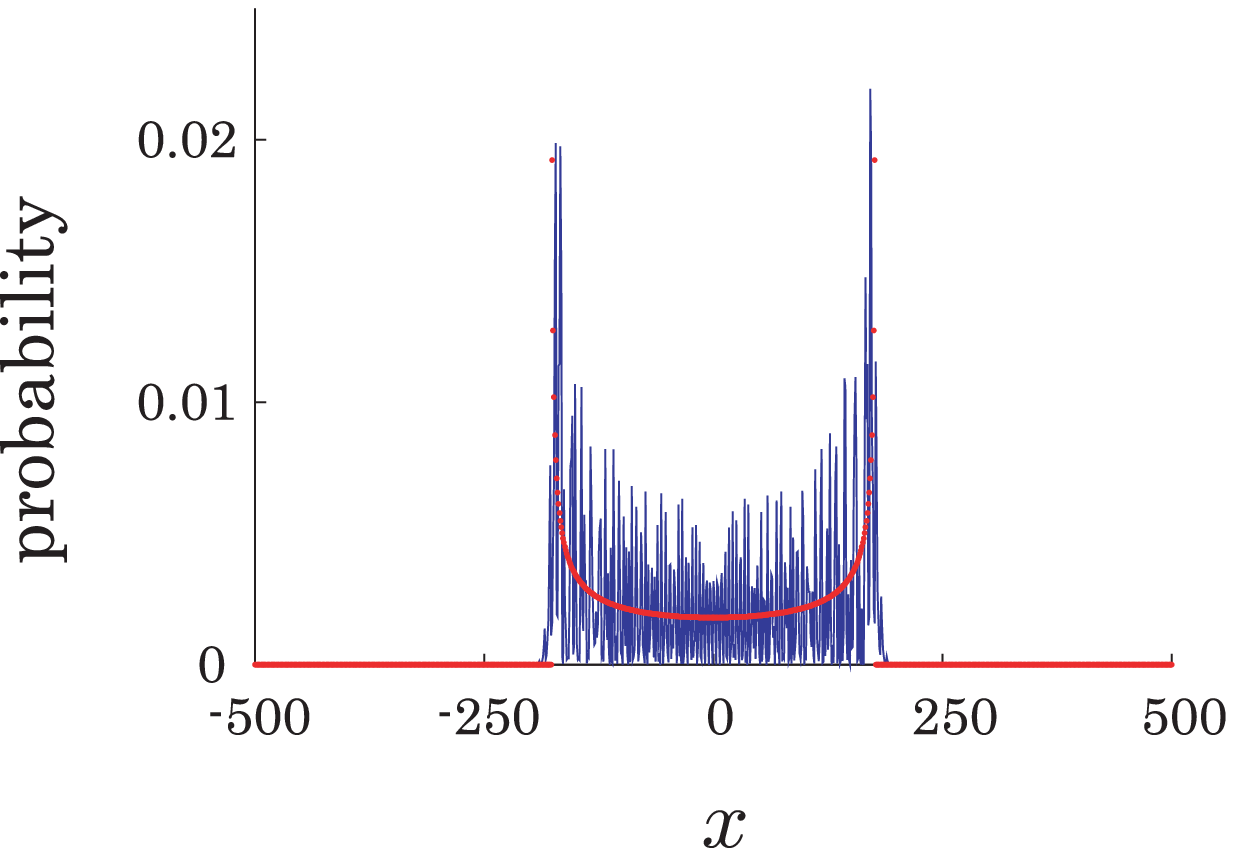}\\[2mm]
  (b) $\gamma_0=1/2\sqrt{2},\, \gamma_1=1/4\sqrt{2}$
  \end{center}
 \end{minipage}
\caption{(Color figure online) The blue lines represent the probability distribution $\mathbb{P}(X_t=x)$ at time $t=500$ and the red points represent the right side of Eq.~\eqref{eq:approximation} as $t=500$. The limit density function approximately reproduces the probability distribution as time $t$ becomes large enough.}
\label{fig:2}
\end{center}
\end{figure}

The limit theorem gives us an approximation of the probability as time $t$ increases large enough,
\begin{equation}
 \mathbb{P}(X_t=x)
 \sim \frac{1}{\pi\sqrt{4\gamma_\xi^2t^2-x^2}}I_{(-2|\gamma_\xi|t, 2|\gamma_\xi|t)}(x)\quad (t\to\infty).\label{eq:approximation}
\end{equation}
Figure~\ref{fig:2} numerically examples the approximation in Eq.~\eqref{eq:approximation}.
Although we see the asymmetry of the probability distribution $\mathbb{P}(X_t=x)$ in the pictures, the approximation is symmetric regarding to $x=0$.

The proof of Theorem~\ref{th:limit} can be demonstrated by Fourier analysis.
The method has been used for finding long-time limit distributions of quantum walks since it was applied to discrete-time quantum walks by Grimmett et. al.~\cite{GrimmettJansonScudo2004}.
Introducing a vector
\begin{align}
 \ket{\hat\psi_t(k)}
 =&
 \begin{bmatrix}
  \hat\psi_{0,t}(k)\\ \hat\psi_{1,t}(k)
 \end{bmatrix}\nonumber\\
 =&
 \,\frac{1}{2I(k)}\,e^{i\sqrt{g(k)}\cdot t}
 \begin{bmatrix}
  I(k)\\ -|I(k)|
 \end{bmatrix}
 +\frac{1}{2I(k)}\,e^{-i\sqrt{g(k)}\cdot t}
 \begin{bmatrix}
  I(k)\\ |I(k)|
 \end{bmatrix},
\end{align}
we organize the $r$-th moments $\mathbb{E}[X_t^r]\, (r=0,1,2,\ldots)$ in the order of time $t$ and focus on the highest order.
With a differential operation $D=i\cdot d/dk$, since we have
\begin{align}
 D^r \ket{\hat\psi_t(k)}
 =&
 \,t^r\cdot e^{i\sqrt{g(k)}\cdot t}\left(-\frac{d}{dk}\sqrt{g(k)}\right)^r\,\frac{1}{2I(k)}
 \begin{bmatrix}
  I(k)\\ -|I(k)|
 \end{bmatrix}\nonumber\\
 & +\,t^r\cdot e^{-i\sqrt{g(k)}\cdot t}\left(\frac{d}{dk}\sqrt{g(k)}\right)^r\,\frac{1}{2I(k)}
 \begin{bmatrix}
  I(k)\\ |I(k)|
 \end{bmatrix}
 +O(t^{r-1}),
\end{align}
the $r$-th moments are arranged in the following forms, 
\begin{align}
 \mathbb{E}[X_t^r]
 =& \sum_{n\in\mathbb{Z}}\, (2n)^r\, \mathbb{P}(X_t=2n) + \sum_{n\in\mathbb{Z}}\, (2n+1)^r\, \mathbb{P}(X_t=2n+1)\nonumber\\
 =& \int_{-\pi}^{\pi}\overline{\hat\psi_{0,t}(k)}\Bigl(D^r\hat\psi_{0,t}(k)\Bigr)\,\frac{dk}{2\pi}\,+\,\int_{-\pi}^{\pi}\overline{\hat\psi_{1,t}(k)}\Bigl(D^r\hat\psi_{1,t}(k)\Bigr)\,\frac{dk}{2\pi}\nonumber\\
 =& \int_{-\pi}^{\pi}\bra{\hat\psi_t(k)}\Bigl(D^r\ket{\hat\psi_t(k)}\Bigr)\,\frac{dk}{2\pi}\nonumber\\
 =& \,t^r\,\left\{\int_{-\pi}^{\pi} \left(\frac{d}{dk}\sqrt{g(k)}\right)^r \,\frac{dk}{4\pi}+\int_{-\pi}^{\pi} \left(-\frac{d}{dk}\sqrt{g(k)}\right)^r \,\frac{dk}{4\pi}\right\}+O(t^{r-1}).
\end{align}
One can derive the limits of the $r$-th moments $\mathbb{E}[(X_t/t)^r]$ as $t\to\infty$,
\begin{equation}
 \lim_{t\to\infty}\mathbb{E}\left[\left(\frac{X_t}{t}\right)^r\right]
  =\int_{-\pi}^{\pi} \left(-\frac{d}{dk}\sqrt{g(k)}\right)^r \,\frac{dk}{4\pi}+\int_{-\pi}^{\pi} \left(\frac{d}{dk}\sqrt{g(k)}\right)^r \,\frac{dk}{4\pi},
\end{equation}
where
\begin{equation}
 \frac{d}{dk}\sqrt{g(k)}=-\frac{2\gamma_0\gamma_1\sin 2k}{\sqrt{\gamma_0^2+\gamma_1^2+2\gamma_0\gamma_1\cos 2k}}.
\end{equation}
With $h(k)=-\frac{d}{dk}\sqrt{g(k)}$, we shrink the integral interval,
\begin{align}
 \int_{-\pi}^{\pi} \bigl(\pm h(k)\bigr)^r\,dk
 =& \int_{-\pi}^{0}\bigl(\pm h(k)\bigr)^r\,dk + \int_{0}^{\pi}\bigl(\pm h(k)\bigr)^r\,dk\nonumber\\
 =& \int_{\pi}^{0} \bigl(\pm h(-k)\bigr)^r\,d(-k) + \int_{0}^{\pi} \bigl(\pm h(k)\bigr)^r\,dk\nonumber\\
 =& \int_{0}^{\pi} \bigl(\mp h(k)\bigr)^r\,dk + \int_{0}^{\pi} \bigl(\pm h(k)\bigr)^r\,dk,
\end{align}
\begin{align}
 \int_{0}^{\pi} \bigl(\pm h(k)\bigr)^r\,dk
 =& \int_{0}^{\frac{\pi}{2}}\bigl(\pm h(k)\bigr)^r\,dk + \int_{\frac{\pi}{2}}^{\pi}\bigl(\pm h(k)\bigr)^r\,dk\nonumber\\
 =& \int_{0}^{\frac{\pi}{2}} \bigl(\pm h(k)\bigr)^r\,dk + \int_{\frac{\pi}{2}}^{0} \bigl(\pm h(\pi-k)\bigr)^r\,d(\pi-k)\nonumber\\
 =& \int_{0}^{\frac{\pi}{2}} \bigl(\pm h(k)\bigr)^r\,dk + \int_{0}^{\frac{\pi}{2}} \bigl(\mp h(k)\bigr)^r\,dk,
\end{align}
resulting in
\begin{equation}
 \int_{-\pi}^{\pi} \bigl(\pm h(k)\bigr)^r\,dk = 2\left\{\int_{0}^{\frac{\pi}{2}} \bigl(\pm h(k)\bigr)^r\,dk + \int_{0}^{\frac{\pi}{2}} \bigl(\mp h(k)\bigr)^r\,dk\right\}.
\end{equation}
Again, the $r$-th moments converge to integral forms,
\begin{equation}
 \lim_{t\to\infty}\mathbb{E}\left[\left(\frac{X_t}{t}\right)^r\right]= \frac{1}{\pi}\left\{\int_{0}^{\frac{\pi}{2}}h(k)^r\,dk + \int_{0}^{\frac{\pi}{2}}\bigl(-h(k)\bigr)^r\,dk\right\},\label{eq:limit_r-th_moments}
\end{equation}
where
\begin{equation}
 h(k)=\frac{2\gamma_0\gamma_1\sin 2k}{\sqrt{\gamma_0^2+\gamma_1^2+2\gamma_0\gamma_1\cos 2k}}.
\end{equation}

\begin{figure}[h]
 \begin{center}
  \begin{minipage}{50mm}
   \begin{center}
    \includegraphics[scale=0.4]{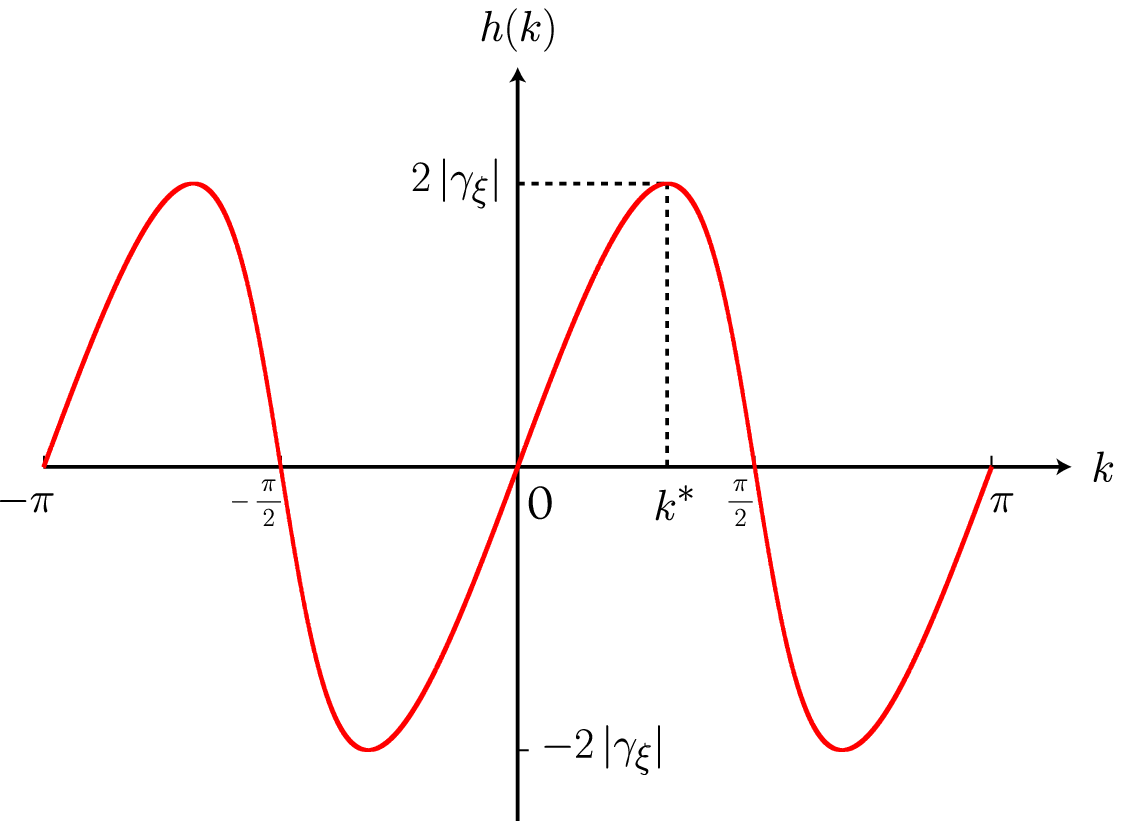}\\
    {(a) $\gamma_0\gamma_1 > 0$}
   \end{center}
  \end{minipage}
  \begin{minipage}{50mm}
   \begin{center}
    \includegraphics[scale=0.4]{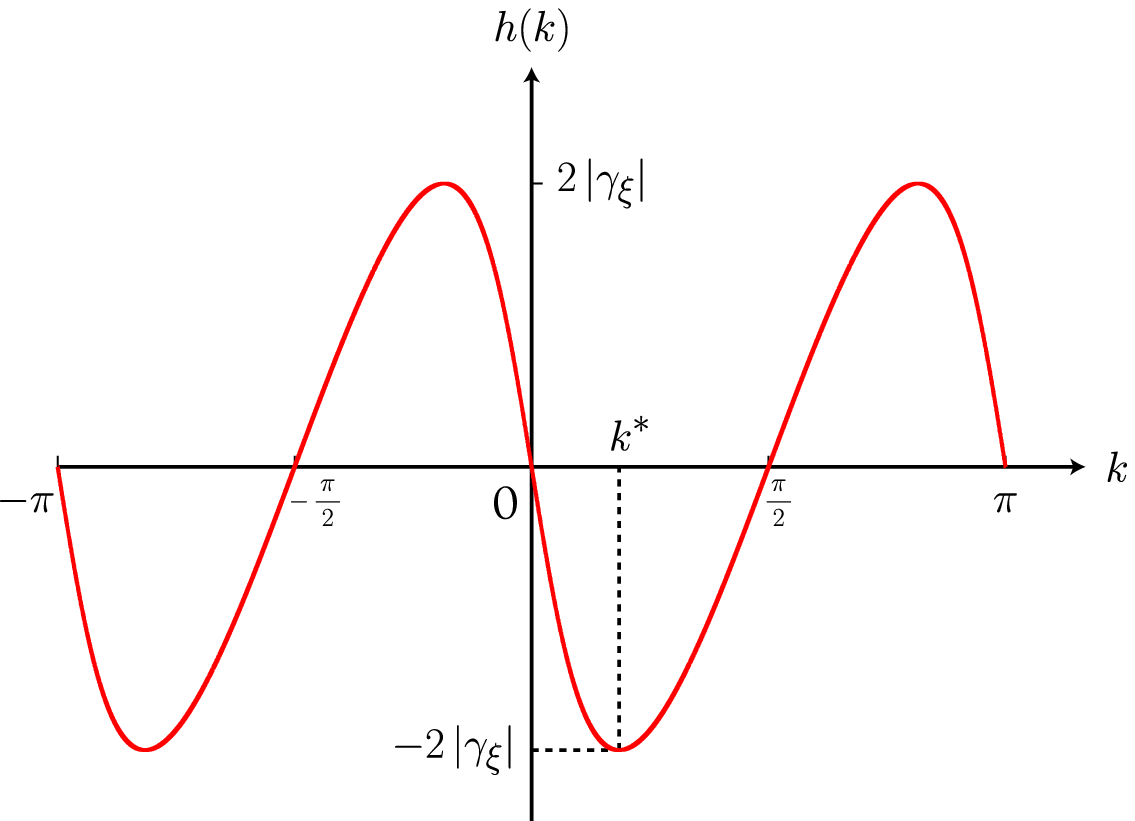}\\
    {(b) $\gamma_0\gamma_1 < 0$}
   \end{center}
  \end{minipage}
  \caption{(Color figure online) The function $h(k)$ is periodic with period $\pi$ and its range is determined by only one of the parameters $\gamma_0$ and $\gamma_1$, that is, $\bigl[-2\,|\gamma_\xi|\,, 2\,|\gamma_\xi|\,\bigr]$.}
  \label{fig:3}
 \end{center}
\end{figure}

Finding the derivative of $h(k)$,
\begin{equation}
 h'(k)=\frac{d}{dk}h(k)=\frac{4\gamma_0^2\gamma_1^2}{g(k)\sqrt{g(k)}}\left(\cos 2k +\frac{\gamma_1}{\gamma_0}\right)\left(\cos 2k +\frac{\gamma_0}{\gamma_1}\right),
\end{equation}
we realize that the function $h(k)$ has a unique extreme value on the interval $[0,\pi/2]$ at a point $k=k^{\ast}\,(\in [0,\pi/2])$,
\begin{equation}
 k^{\ast}=\left\{\begin{array}{ll}
	   \displaystyle \frac{1}{2}\arccos\left(-\frac{\gamma_1}{\gamma_0}\right)& (\,|\gamma_0|>|\gamma_1|\,)\\[5mm]
	   \displaystyle \frac{1}{2}\arccos\left(-\frac{\gamma_0}{\gamma_1}\right)& (\,|\gamma_0|<|\gamma_1|\,)
		 \end{array}\right.,
\end{equation}
\begin{align}
 h(k^{\ast})=& \left\{\begin{array}{ll}
		\displaystyle 2\,|\gamma_1| \cdot \frac{\gamma_0\gamma_1}{\,|\gamma_0\gamma_1|\,} & (\,|\gamma_0|>|\gamma_1|\,)\\[5mm]
		       \displaystyle 2\,|\gamma_0| \cdot \frac{\gamma_0\gamma_1}{\,|\gamma_0\gamma_1|\,} & (\,|\gamma_0|<|\gamma_1|\,)
		      \end{array}\right.\nonumber\\[1mm]
 =& 2\,|\gamma_\xi| \cdot \frac{\gamma_0\gamma_1}{\,|\gamma_0\gamma_1|\,}\nonumber\\[1mm]
 =& \left\{\begin{array}{rl}
    2\,|\gamma_\xi| & \quad (\gamma_0\gamma_1 > 0)\\
	    -2\,|\gamma_\xi| & \quad (\gamma_0\gamma_1 <0)
	   \end{array}\right.,
\end{align}
where $\xi\in\left\{0,1\right\}$ is the subscription such that $|\gamma_\xi|=\min\left\{\,|\gamma_0|,|\gamma_1|\,\right\}$.
We also see the sign of the derivative,
\begin{align}
  \left\{\begin{array}{lll}
   h'(k)>0 & (k\in [0,k^{\ast}))\\
   h'(k)<0 & (k\in (k^{\ast},\pi/2])\\
	 \end{array}\right.
  \quad (\gamma_0\gamma_1>0),\\[1mm]
  \left\{\begin{array}{lll}
   h'(k)<0 & (k\in [0,k^{\ast}))\\
   h'(k)>0 & (k\in (k^{\ast},\pi/2])\\
	 \end{array}\right.
  \quad (\gamma_0\gamma_1<0).
\end{align}
The function $h(k)$ is a periodic function with period $\pi$ and holds $h(\pi-k)=-h(k)$.
The extreme value $h(k^{\ast})$ is equal to the maximum (resp. minimum) value of function $h(k)$ if the value of $\gamma_0\gamma_1$ is positive (resp. negative).
Figure~\ref{fig:3} visualizes the function $h(k)$ on the interval $[-\pi,\pi]$.

\begin{figure}[h]
 \begin{center}
  \begin{minipage}{50mm}
   \begin{center}
    \includegraphics[scale=0.4]{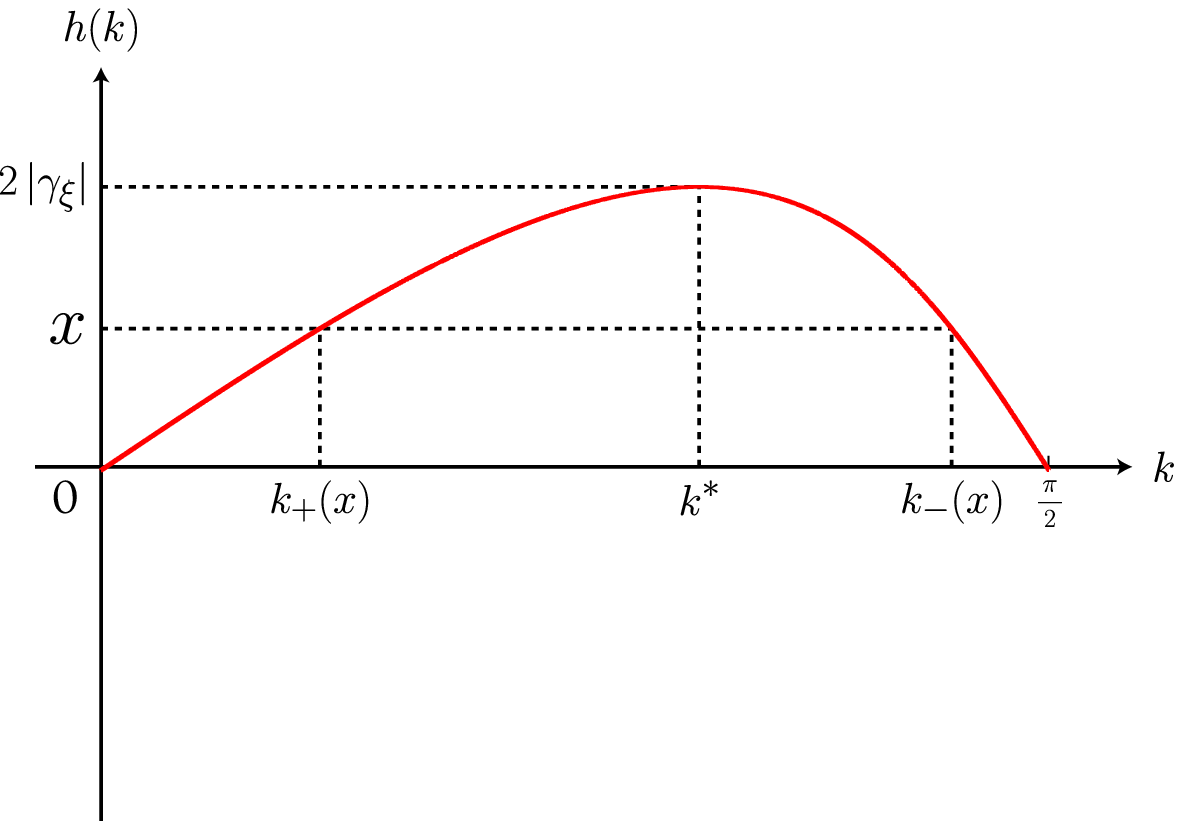}\\
    {(a) $\gamma_0\gamma_1 > 0$}
   \end{center}
  \end{minipage}
  \begin{minipage}{50mm}
   \begin{center}
    \includegraphics[scale=0.4]{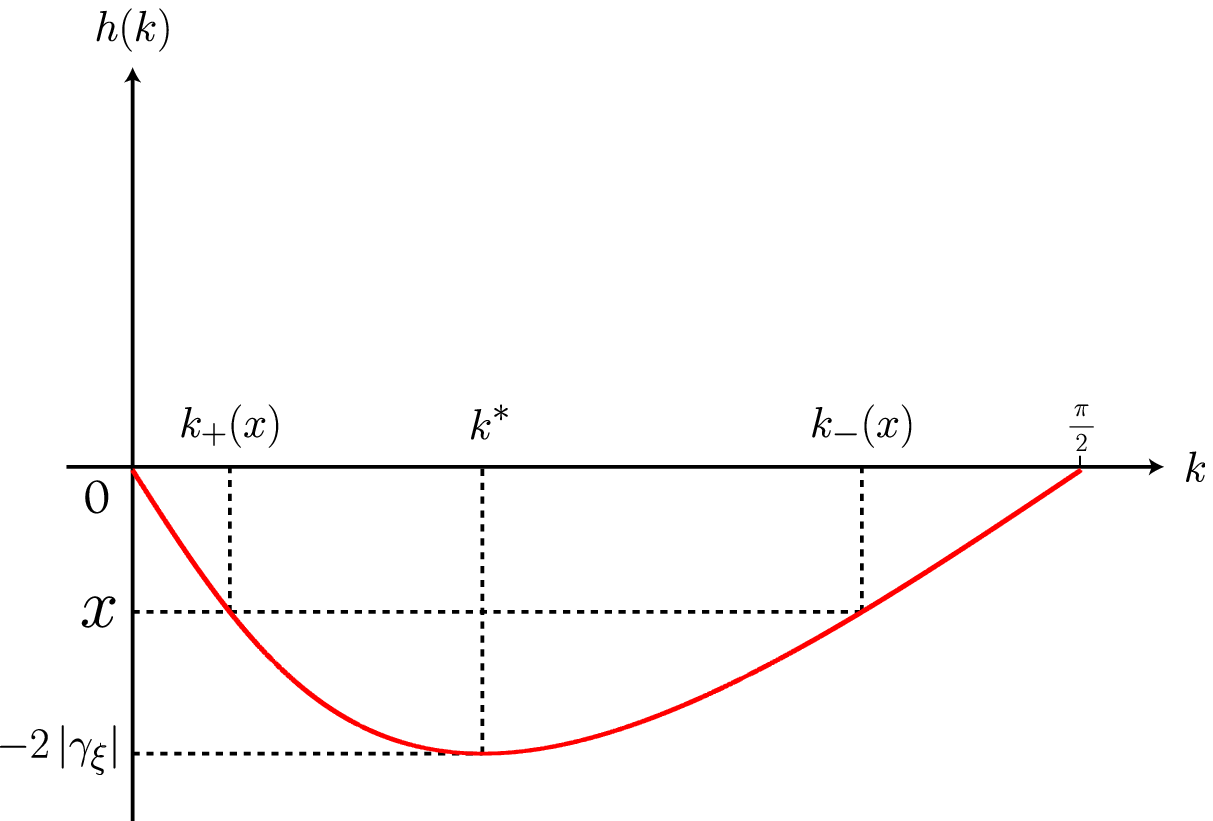}\\
    {(b) $\gamma_0\gamma_1 < 0$}
   \end{center}
  \end{minipage}
  \caption{(Color figure online) The functions $k_{\pm}(x)\,(\in [0,\pi/2]\,)$ satisfy $h(k_{\pm}(x))=x$.}
  \label{fig:4}
 \end{center}
\end{figure}

Substituting $h(k)=x$, we get another representation of the integrals,
\begin{align}
 \int_{0}^{\frac{\pi}{2}}\Bigl(\pm h(k)\Bigr)^r\,dk
 =& \int_{0}^{k^{\ast}}\Bigl(\pm h(k)\Bigr)^r\,dk + \int_{k^{\ast}}^{\frac{\pi}{2}}\Bigl(\pm h(k)\Bigr)^r\,dk\nonumber\\
 =& \int_{0}^{h(k^{\ast})}(\pm x)^r \frac{dk_{+}(x)}{dx}\,dx + \int_{h(k^{\ast})}^{0}(\pm x)^r \frac{dk_{-}(x)}{dx}\,dx\nonumber\\
 =& \int_{0}^{h(k^{\ast})}(\pm x)^r \biggl(\frac{dk_{+}(x)}{dx}-\frac{dk_{-}(x)}{dx}\biggr)\,dx,
\end{align}
where the functions $k_{\pm}(x)\,(\in [0,\pi/2]\,)$ satisfy $h(k_{\pm}(x))=x$ (See also Fig.~\ref{fig:4}.), that is,
\begin{equation}
 k_{\pm}(x)=
  \left\{\begin{array}{ll}
   \displaystyle \frac{1}{2}\arccos\left(\frac{-x^2\pm\sqrt{4\gamma_0^2-x^2}\sqrt{4\gamma_1^2-x^2}}{4\gamma_0\gamma_1}\right) & (\gamma_0\gamma_1>0)\\[5mm]
   \displaystyle \frac{1}{2}\arccos\left(\frac{-x^2\mp\sqrt{4\gamma_0^2-x^2}\sqrt{4\gamma_1^2-x^2}}{4\gamma_0\gamma_1}\right) & (\gamma_0\gamma_1<0)
	 \end{array}\right.,
\end{equation}
whose derivatives are computed and organized,
\begin{align}
 & \frac{d}{dx}\left(\frac{1}{2}\arccos\left(\frac{-x^2\pm\sqrt{4\gamma_0^2-x^2}\sqrt{4\gamma_1^2-x^2}}{4\gamma_0\gamma_1}\right)\right)\nonumber\\
 =& \pm \frac{1}{2}\cdot\frac{x}{\,|x|\,}\cdot\frac{\,|\gamma_0\gamma_1|\,}{\gamma_0\gamma_1}\cdot\frac{\,\Bigl|\,\sqrt{4\gamma_0^2-x^2}\pm\sqrt{4\gamma_1^2-x^2}\,\Bigr|\,}{\sqrt{4\gamma_0^2-x^2}\sqrt{4\gamma_1^2-x^2}}\nonumber\\
 =:&\, f_{\pm}(x).
\end{align}

\noindent
Note that the equation $h(k)=x\,(k\in [0,\pi/2]\,)$ has two solutions under the condition $x\in [\,0,2\,|\gamma_\xi|\,]\, (\gamma_0\gamma_1>0)$\,/\,$x\in [\,-2\,|\gamma_\xi|,0\,]\, (\gamma_0\gamma_1<0)$, and the functions $k_{\pm}(x)$ are the solutions where they hold $k_{+}(x) \leq k_{-}(x)$.

Let us wrap up the computation,
\begin{align}
 &\int_{0}^{\frac{\pi}{2}} h(k)^r\,dk + \int_{0}^{\frac{\pi}{2}} \bigl(-h(k)\bigr)^r\,dk\nonumber\\
=& \left\{\begin{array}{r}
    \displaystyle \int_{0}^{h(k^{\ast})}x^r\,\bigl(f_{+}(x)-f_{-}(x)\bigr)\,dx + \int_{h(k^{\ast})}^{0}(-x)^r\,\Bigl\{-\bigl(f_{+}(x)-f_{-}(x)\bigr)\Bigr\}\,dx\\
	   (\gamma_0\gamma_1>0)\\[3mm]
	   \displaystyle \int_{0}^{h(k^{\ast})}x^r\,\bigl(f_{-}(x)-f_{+}(x)\bigr)\,dx + \int_{h(k^{\ast})}^{0}(-x)^r\,\Bigl\{-\bigl(f_{-}(x)-f_{+}(x)\bigr)\Bigr\}\,dx\\
	   (\gamma_0\gamma_1<0)
	  \end{array}\right.\nonumber\\
=& \left\{\begin{array}{r}
    \displaystyle \int_{0}^{h(k^{\ast})}x^r\,\bigl(f_{+}(x)-f_{-}(x)\bigr)\,dx + \int_{-h(k^{\ast})}^{0}x^r\,\Bigl\{-\bigl(f_{+}(-x)-f_{-}(-x)\bigr)\Bigr\}\,d(-x)\\
	   (\gamma_0\gamma_1>0)\\[3mm]
	   \displaystyle \int_{0}^{h(k^{\ast})}x^r\,\bigl(f_{-}(x)-f_{+}(x)\bigr)\,dx + \int_{-h(k^{\ast})}^{0}x^r\,\Bigl\{-\bigl(f_{-}(-x)-f_{+}(-x)\bigr)\Bigr\}\,d(-x)\\
	   (\gamma_0\gamma_1<0)
	  \end{array}\right.\nonumber\\
=& \left\{\begin{array}{l}
    \displaystyle \int_{0}^{2\,|\gamma_\xi|}x^r\,\frac{1}{\sqrt{4\gamma_\xi^2-x^2}}\,dx + \int_{-2\,|\gamma_\xi|}^{0}x^r\,\frac{1}{\sqrt{4\gamma_\xi^2-x^2}}\,dx\\
	   \hspace{\fill} (\gamma_0\gamma_1>0)\\[3mm]
	   \displaystyle \int_{0}^{-2\,|\gamma_\xi|}x^r\,\left(-\,\frac{1}{\sqrt{4\gamma_\xi^2-x^2}}\right)\,dx + \int_{2\,|\gamma_\xi|}^{0}x^r\,\left(-\,\frac{1}{\sqrt{4\gamma_\xi^2-x^2}}\right)\,dx\\
	   \hspace{\fill} (\gamma_0\gamma_1<0)
	  \end{array}\right.\nonumber\\
 =& \int_{-2\,|\gamma_\xi|}^{2\,|\gamma_\xi|}x^r\,\frac{1}{\sqrt{4\gamma_\xi^2-x^2}}\,dx,
\end{align}
in which we have used
\begin{align}
 & f_{+}(x)-f_{-}(x)\nonumber\\
=& \,\frac{1}{2}\cdot\frac{x}{\,|x|\,}\cdot\frac{\,|\gamma_0\gamma_1|\,}{\gamma_0\gamma_1}\cdot\frac{\,\Bigl|\,\sqrt{4\gamma_0^2-x^2}+\sqrt{4\gamma_1^2-x^2}\,\Bigr|\,+\,\Bigl|\sqrt{4\gamma_0^2-x^2}-\sqrt{4\gamma_1^2-x^2}\,\Bigr|\,}{\sqrt{4\gamma_0^2-x^2}\sqrt{4\gamma_1^2-x^2}}\nonumber\\[3mm]
=& \left\{\begin{array}{ll}
    \displaystyle \frac{x}{\,|x|\,}\cdot\frac{\,|\gamma_0\gamma_1|\,}{\gamma_0\gamma_1}\cdot\frac{1}{\sqrt{4\gamma_1^2-x^2}} & \quad (\,|\gamma_0|>|\gamma_1|\,)\\[5mm]
    \displaystyle \frac{x}{\,|x|\,}\cdot\frac{\,|\gamma_0\gamma_1|\,}{\gamma_0\gamma_1}\cdot\frac{1}{\sqrt{4\gamma_0^2-x^2}} & \quad (\,|\gamma_0|<|\gamma_1|\,)
	  \end{array}\right.\nonumber\\[3mm]
=&\frac{x}{\,|x|\,}\cdot\frac{\,|\gamma_0\gamma_1|\,}{\gamma_0\gamma_1}\cdot\frac{1}{\sqrt{4\gamma_\xi^2-x^2}}.
\end{align}
Reminding Eq.~\eqref{eq:limit_r-th_moments}, we find another representation for the limits of the $r$-th moments,
\begin{align}
 \lim_{t\to\infty}\mathbb{E}\left[\left(\frac{X_t}{t}\right)^r\right]
 =& \int_{-2\,|\gamma_\xi|}^{2\,|\gamma_\xi|}x^r\,\frac{1}{\pi\sqrt{4\gamma_\xi^2-x^2}}\,dx\nonumber\\
 =& \int_{-\infty}^{\infty}x^r\,\frac{1}{\pi\sqrt{4\gamma_\xi^2-x^2}}\,I_{(\,-2\,|\gamma_\xi|, 2\,|\gamma_\xi|\,)}\,dx,\label{eq:limit_r-th_moments_2}
\end{align}
where
\begin{equation}
 I_{(\,-2\,|\gamma_\xi|, 2\,|\gamma_\xi|\,)}(x)
  =\left\{\begin{array}{cl}
    1&(\,-2\,|\gamma_\xi| < x < 2\,|\gamma_\xi|\,)\\
	   0&(\mbox{otherwise})
	  \end{array}\right..
\end{equation}
Equation~\eqref{eq:limit_r-th_moments_2} guarantees the statement of Theorem~\ref{th:limit} and its proof has been completed.

\section{Summary}
\label{sec:summary}
We studied a continuous-time quantum walk on $\mathbb{Z}=\left\{0,\pm1,\pm2,\ldots\right\}$ and its Hamiltonian was given by a spatially 2-periodic matrix.
The probability amplitude $\left\{\psi_t(x) : x\in\mathbb{Z}\right\}$ was found in integral forms by Fourier analysis and we observed asymmetric probability distributions.
We also discovered a limit distribution of the scaled position $X_t/t$ as $t\to\infty$, which produced an approximation to the probability distribution $\mathbb{P}(X_t=x)$ as time $t$ was large enough.
Differently from the probability distribution, the limit density function was symmetric for any values of $\gamma_0$ and $\gamma_1$.
It was completely determined by only one of the parameters $\gamma_0$ and $\gamma_1$, represented by $\gamma_\xi$ such that $|\gamma_\xi|=\min\left\{\,|\gamma_0|,\, |\gamma_1|\,\right\}$.

A similar property was also reported for discrete-time quantum walks in the past studies~\cite{MachidaKonno2010,MachidaGrunbaum2018}.
The systems of discrete-time quantum walks got updated by two unitary matrices which alternately operated the quantum walker.
Such a quantum walk is called a 2-period time-dependent quantum walk.
Machida and Konno~\cite{MachidaKonno2010} studied a long-time limit theorem for a 2-state quantum walk on $\mathbb{Z}$.
The limit distribution was determined by only one of the two unitary matrices in a case, that is, the other did not work on the limit distribution at all.
Machida and Gr\"{u}nbaum~\cite{MachidaGrunbaum2018} derived a limit distribution of a 4-state quantum walk on $\mathbb{Z}$ and it held the same property as the 2-state quantum walk.

Although two parameters operated the continuous-time quantum walk, one parameter did not affect the quantum walker in approximation as time $t$ increased enough.
It would be a future challenge to use such an interesting discovery for applications modeled by Schr\"{o}dinger equations.
\bigskip\bigskip

\begin{center}
{\bf Acknowledgements}
\end{center}
The author is supported by JSPS Grant-in-Aid for Scientific Research (C) (No. 23K03220).


\clearpage

\appendix
\section{$|\gamma_0|=|\gamma_1|$}
We briefly see the continuous-time quantum walk in the case of $|\gamma_0|=|\gamma_1|$ (i.e. $\gamma_1=\pm\gamma_0$).
Since one can derive the results, which we will find here, in a similar way as shown in Sects.~\ref{sec:probability_amplitude} and \ref{sec:limit_distribution}, the proofs are all omitted in this section.
The reason we separately observe the case of $|\gamma_0|=|\gamma_1|$ is that the denominator of the function
\begin{equation}
 h(k)=\frac{2\gamma_0\gamma_1\sin 2k}{\sqrt{\gamma_0^2+\gamma_1^2+2\gamma_0\gamma_1\cos 2k}}\quad (k\in [-\pi, \pi)),
\end{equation}
becomes zero at $k=\pm\pi/2$ (resp. $k=-\pi, 0$) if $\gamma_1=\gamma_0$ (resp. $\gamma_1=-\gamma_0$).
But, we will realize that the results can be contained in the representations which we got in the case of $|\gamma_0|\neq |\gamma_1|$.

With some notations
\begin{align}
 & \gamma=\gamma_0,\\
 & n=0,1,2,\ldots,\\
 & J_n(x) : \mbox{Bessel functions of the first kind},
\end{align}
one can find the probability amplitude and the limit theorem as follows.

\begin{enumerate}
 \item Case : $\gamma_1=\gamma_0\, (=\,\gamma)$
       \begin{align}
	\psi_t(2n)
	=& \int_{0}^{\frac{\pi}{2}}\frac{2}{\pi}\cos \bigl(\,2nk\bigr) \cos(2\gamma\, t\cos k)\,dk\nonumber\\
	=& \int_{0}^{\frac{\pi}{2}}\frac{2}{\pi}\cos \bigl(\,2nk\bigr) \cos(2\,|\gamma|\, t\cos k)\,dk\nonumber\\
	=& (-1)^n J_{2n}(2\,|\gamma|\,t)\nonumber\\
	=&\,i^{2n} J_{2n}(2\,|\gamma|\,t),\\[3mm]
	\psi_t(-2n)
	=& \int_{0}^{\frac{\pi}{2}}\frac{2}{\pi}\cos \bigl(\,-2nk\bigr) \cos(2\gamma\, t\cos k)\,dk\nonumber\\
	=& \int_{0}^{\frac{\pi}{2}}\frac{2}{\pi}\cos \bigl(\,2nk\bigr) \cos(2\,|\gamma|\, t\cos k)\,dk\nonumber\\
	=& (-1)^n J_{2n}(2\,|\gamma|\,t)\nonumber\\
	=&\,i^{2n} J_{2n}(2\,|\gamma|\,t),\\[3mm]
	\psi_t(2n+1)
	=& -i\,\int_{0}^{\frac{\pi}{2}}\frac{2}{\pi}\cos \bigl(\,(2n+1)k\bigr) \sin(2\gamma\, t\cos k)\,dk\nonumber\\
	=& -i\,\frac{\gamma}{\,|\gamma|\,}\int_{0}^{\frac{\pi}{2}}\frac{2}{\pi}\cos \bigl(\,(2n+1)k\bigr) \sin(2\,|\gamma|\, t\cos k)\,dk\nonumber\\
	=& -(-1)^n\, i\,\frac{\gamma}{\,|\gamma|\,}J_{2n+1}(2\,|\gamma|\,t)\nonumber\\
	=& -i^{2n+1}\,\frac{\gamma}{\,|\gamma|\,}J_{2n+1}(2\,|\gamma|\,t),\label{eq:case1-3}\\[3mm]
	\psi_t(-2n-1)
	=& -i\,\int_{0}^{\frac{\pi}{2}}\frac{2}{\pi}\cos \bigl(\,(-2n-1)k\bigr) \sin(2\gamma\, t\cos k)\,dk\nonumber\\
	=&  -i\,\frac{\gamma}{\,|\gamma|\,}\int_{0}^{\frac{\pi}{2}}\frac{2}{\pi}\cos \bigl(\,(2n+1)k\bigr) \sin(2\,|\gamma|\, t\cos k)\,dk\nonumber\\
	=& -(-1)^n\, i\,\frac{\gamma}{\,|\gamma|\,}J_{2n+1}(2\,|\gamma|\,t)\nonumber\\
	=& -i^{2n+1}\,\frac{\gamma}{\,|\gamma|\,}J_{2n+1}(2\,|\gamma|\,t).\label{eq:case1-4}
       \end{align}
       These representations of the amplitude match Corollary~1 in Konno~\cite{Konno2005b} as $\gamma=-\frac{1}{2}$.

 \item Case : $\gamma_1=-\gamma_0\, (=-\gamma)$
       \begin{align}
	\psi_t(2n)
	=& (-1)^n\int_{0}^{\frac{\pi}{2}}\frac{2}{\pi}\cos \bigl(\,2nk\bigr) \cos(2\gamma\, t\cos k)\,dk\nonumber\\
	=& (-1)^n\int_{0}^{\frac{\pi}{2}}\frac{2}{\pi}\cos \bigl(\,2nk\bigr) \cos(2\,|\gamma|\, t\cos k)\,dk\nonumber\\
	=& J_{2n}(2\,|\gamma|\,t),\\[3mm]
	\psi_t(-2n)
	=& (-1)^{-n}\int_{0}^{\frac{\pi}{2}}\frac{2}{\pi}\cos \bigl(\,-2nk\bigr) \cos(2\gamma\, t\cos k)\,dk\nonumber\\
	=& (-1)^n\int_{0}^{\frac{\pi}{2}}\frac{2}{\pi}\cos \bigl(\,2nk\bigr) \cos(2\,|\gamma|\, t\cos k)\,dk\nonumber\\
	=& J_{2n}(2\,|\gamma|\,t),\\[3mm]
	\psi_t(2n+1)
	=& -(-1)^n\, i\,\int_{0}^{\frac{\pi}{2}}\frac{2}{\pi}\cos \bigl(\,(2n+1)k\bigr) \sin(2\gamma\, t\cos k)\,dk\nonumber\\
	=& -(-1)^n\, i\,\frac{\gamma}{\,|\gamma|\,}\int_{0}^{\frac{\pi}{2}}\frac{2}{\pi}\cos \bigl(\,(2n+1)k\bigr) \sin(2\,|\gamma|\, t\cos k)\,dk\nonumber\\
	=& -i\,\frac{\gamma}{\,|\gamma|\,}J_{2n+1}(2\,|\gamma|\,t),\label{eq:case2-3}\\[3mm]
	\psi_t(-2n-1)
	=& -(-1)^{-n-1}\,i\,\int_{0}^{\frac{\pi}{2}}\frac{2}{\pi}\cos \bigl(\,(-2n-1)k\bigr) \sin(2\gamma\, t\cos k)\,dk\nonumber\\
	=&  (-1)^n\, i\,\frac{\gamma}{\,|\gamma|\,}\int_{0}^{\frac{\pi}{2}}\frac{2}{\pi}\cos \bigl(\,(2n+1)k\bigr) \sin(2\,|\gamma|\, t\cos k)\,dk\nonumber\\
	=& \,i\,\frac{\gamma}{\,|\gamma|\,}J_{2n+1}(2\,|\gamma|\,t).\label{eq:case2-4}
       \end{align}
\end{enumerate}

We have expressed $\sin(2\,\gamma\,t\cos k)$ in a different way,
\begin{equation}
 \sin(2\,\gamma\,t\cos k)=\frac{\gamma}{\,|\gamma|\,}\sin(2\,|\gamma|\,t\cos k),
\end{equation}
so that the probability amplitude is described by Bessel functions $J_{2n+1}(2\,|\gamma|\,t)$ in Eqs.~\eqref{eq:case1-3}, \eqref{eq:case1-4}, \eqref{eq:case2-3}, and \eqref{eq:case2-4}.
We should note that these equations are allowed to be contained in the results in the case of $|\gamma_0|\neq |\gamma_1|$.

\begin{figure}[h]
\begin{center}
 \begin{minipage}{50mm}
  \begin{center}
   \includegraphics[scale=0.4]{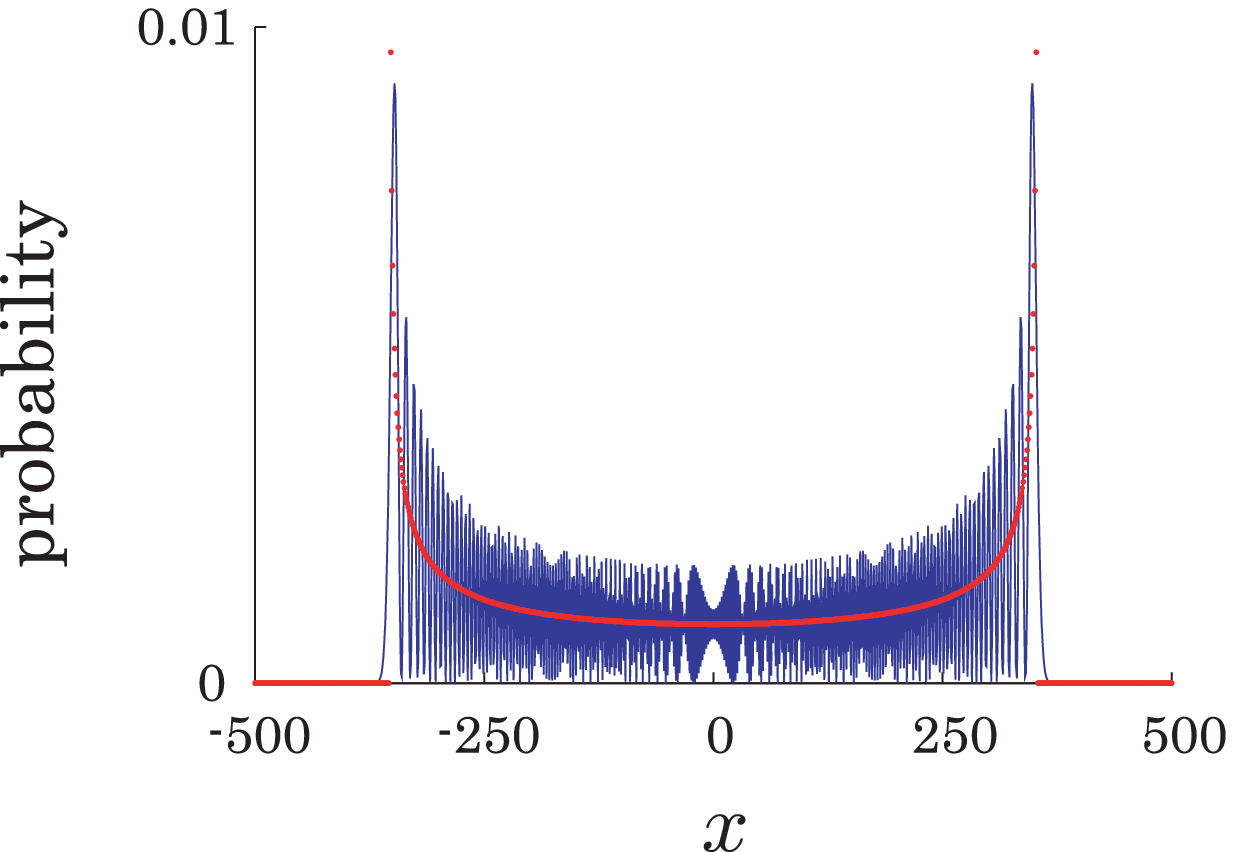}\\[2mm]
  (a) $\gamma_0=\gamma_1=1/2\sqrt{2}$
  \end{center}
 \end{minipage}\hspace{10mm}
 \begin{minipage}{50mm}
  \begin{center}
   \includegraphics[scale=0.4]{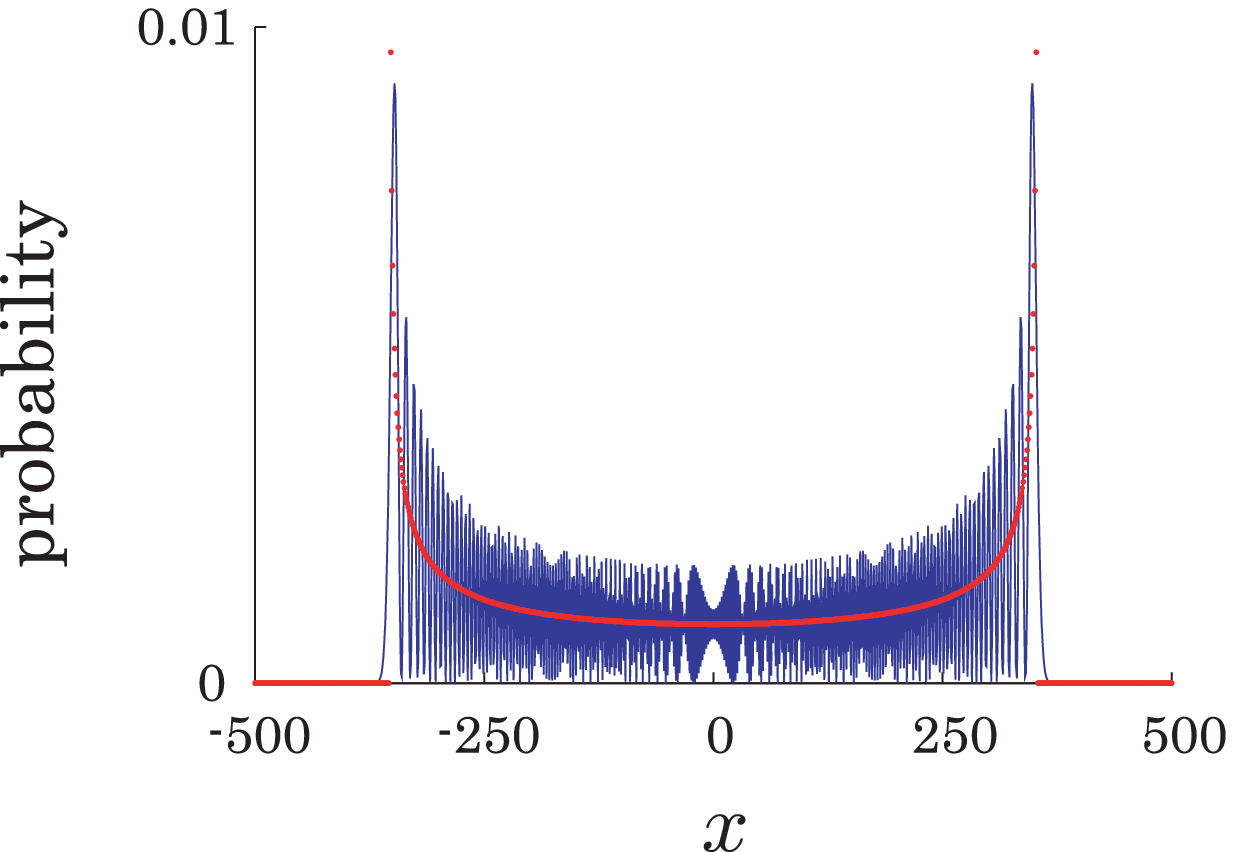}\\[2mm]
  (b) $\gamma_0=-\gamma_1=1/2\sqrt{2}$
  \end{center}
 \end{minipage}
\caption{(Color figure online) The blue lines represent the probability distribution $\mathbb{P}(X_t=x)$ at time $t=500$ and the red points represent the right side of Eq.~\eqref{eq:approximation-a} as $t=500$. The limit density function approximately reproduces the probability distribution as time $t$ becomes large enough.}
\label{fig:5}
\end{center}
\end{figure}

We find the walker at position $x\in\mathbb{Z}$ at time $t$ with probability
\begin{align}
 \mathbb{P}(X_t=x)=J_{|x|}(2\,|\gamma|\,t)^2,
\end{align}
which is equivalent to the probability that the quantum walker, whose position at time $t$ is represented by $Y_t$, is observed at position $x$ at time $2\,|\gamma|\,t$ in the case of $\gamma_0=\gamma_1=-\frac{1}{2}$.
Using Theorem~1 in Konno~\cite{Konno2005b}, for a real number $x$, we have a convergence,
\begin{align}
 \lim_{t\to\infty}\mathbb{P}\left(\frac{X_t}{t}\leq x\right)
 =&\lim_{t\to\infty}\mathbb{P}\left(\frac{Y_{2\,|\gamma|\,t}}{t}\leq x\right)\nonumber\\
 =&\lim_{t\to\infty}\mathbb{P}\left(\frac{Y_{2\,|\gamma|\,t}}{2\,|\gamma|\,t}\leq \frac{x}{2\,|\gamma|\,}\right)\nonumber\\
 =&\int_{-\infty}^{x/2\,|\gamma|}\frac{1}{\pi\sqrt{1-y^2}}\,I_{(\,-1, 1\,)}(y)\,dy\nonumber\\
 =&\int_{-\infty}^{x}\frac{1}{\pi\sqrt{1-\left(\frac{y}{2\,|\gamma|\,}\right)^2}}\,I_{(\,-1, 1\,)}\left(\frac{y}{2\,|\gamma|\,}\right)\,d\left(\frac{y}{2\,|\gamma|\,}\right)\nonumber\\
 =&\int_{-\infty}^{x}\frac{1}{\pi\sqrt{4\gamma^2-y^2}}\,I_{(\,-2\,|\gamma|, 2\,|\gamma|\,)}(y)\,dy,
\end{align}
whose representation is allowed to be contained in Theorem~\ref{th:limit}.
The limit density function produces an approximation to the probability distribution as time $t$ is large enough,
\begin{equation}
 \mathbb{P}(X_t=x)
 \sim \frac{1}{\pi\sqrt{4\gamma^2t^2-x^2}}I_{(-2|\gamma|t, 2|\gamma|t)}(x)\quad (t\to\infty),\label{eq:approximation-a}
\end{equation}
and we confirm the comparison between both sides in Fig.~\ref{fig:5}.
We should note that the probability distribution and the limit density function are both symmetric regarding to $x=0$ in the case of $|\gamma_0|=|\gamma_1|$.

\end{document}